\documentclass[preprint2]{aastex}

\newcommand{\xvec}  {\mbox{\boldmath $x$}}
\newcommand{\yvec}  {\mbox{\boldmath $y$}}

\newcommand{\uvec}  {\mbox{\boldmath $u$}}
\newcommand{\rvec}  {\mbox{\boldmath $r$}}
\newcommand{\lvec}  {\mbox{\boldmath $l$}}

\def\plotfiddle#1#2#3#4#5#6#7{\centering \leavevmode
\vbox to#2{\rule{0pt}{#2}}
\includegraphics{#1}}
\begin{document}

\title{A New Channel to Search for Extra-solar Systems with\\
       Multiple Planets via Gravitational Microlensing}

\author{Cheongho Han}
\affil{Department of Physics, \\
       Chungbuk National University, Chongju 361-763, Korea}
\email{cheongho@astroph.chungbuk.ac.kr}

\bigskip
\author{Myeong-Gu Park}
\affil{Department of Astronomy \& Atmospheric Sciences, \\
       Kyungpook National University, Taegu 702-701, Korea}
\email{mgp@knu.ac.kr}

\begin{abstract}
Microlensing is one of the promising techniques that can be used to search
for extra-solar systems.  Planet detection via microlensing is possible 
because the event caused by a lens system having a planet can produce
noticeable anomalies in the lensing light curve if the source star passes
close to the deviation region induced by the planet.  Gaudi, Naber \& Sackett 
pointed out that if an event is caused by a lens system containing more than 
two planets, all planets will affect the central region of the magnification 
pattern, and thus the existence of the multiple planets can be inferred by 
detecting additionally deformed anomalies from intensive monitoring of high 
magnification events.  Unfortunately, this method has important limitations 
in identifying the existence of multiple planets and determining their 
parameters (the mass ratio and the instantaneous projected separation) due 
to the degeneracy of the resulting light curve anomalies from those induced 
by a single planet and the complexity of multiple planet lensing models.  
In this paper, we propose a new channel to search for multiple planets via 
microlensing.  The method is based on the finding of Han et al.\ that the 
lensing light curve anomalies induced by multiple planets are well 
approximated by the superposition of those of the single planet systems where 
the individual planet-primary pairs act as independent lens systems.  Then, 
if the source trajectory passes both of the outer deviation regions induced 
by the individual planets, one can unambiguously identify the existence of 
the multiple planets.  We illustrate that the probability of successively 
detecting light curve anomalies induced by two Jovian-mass planets located 
in the lensing zone through this channel will be substantial.  Since the 
individual anomalies can be well modeled by much simpler single planet 
lensing models, the proposed method has an important advantage of allowing 
one to accurately determine the parameters of the individual planets.
\end{abstract}

\section{Introduction}

Microlensing is one of the promising techniques that can be used to search 
for extra-solar planets, especially located at large distances 
\citep{perryman00}.  Planet detection via microlensing is possible because 
the lensing event caused by a lens system containing a planet can produce 
noticeable anomalies in the resulting light curve when the source passes 
close to the lens caustics, which represents the source positions on which 
the lensing magnification of a point source event becomes infinity 
\citep{mao91, gould92}.\footnote{See \S\ 2 for more details about the 
caustics.} For a lens system with a planet, there exist two or three 
disconnected sets of caustics.  Among them, one is located close to the 
primary lens (central caustic) and the other(s) is (are) located away from 
the primary lens (planetary caustic[s]).  Accordingly, there exist two types 
of planet-induced anomalies: one affected by the planetary caustic (type I 
anomaly) and the other affected by the central caustic (type II anomaly) 
\citep{han01a}.  Compared to the frequency of type I anomalies, type II 
anomaly occurs with a relatively low frequency due to the smaller size of 
the central caustic compared to the corresponding planetary caustic.  However, 
the efficiency of detecting type II anomalies can be high because intensive 
monitoring is possible due to the predictable time of anomalies, i.e.\ near 
the peak of magnification, and the known type of candidate events for 
intensive follow-up monitoring, i.e.\ very high magnification events 
\citep{griest98}.

Keeping the high efficiency of type II anomaly detections in mind, 
\citet{gaudi98b} pointed out that if an event is caused by a lens system 
having multiple planets located in the lensing zone, within which the chance 
for the occurrence of planet-induced anomalies is maximized\footnote{See 
also \S\ 2 for more details about the lensing zone.}, all planets will 
affect the central region of magnification pattern, and thus the existence 
of multiple planets can be inferred by detecting additionally deformed 
anomalies from intensive monitoring of high magnification events.  This 
method, however, has important limitations in identifying the existence of 
multiple planets and determining their parameters (the mass ratio and the 
instantaneous separation between each planet and the host star).  This is 
because the anomalies induced by multiple planets are in many cases 
qualitatively degenerate from that induced by a single planet and even if 
the existence of the multiple planets is known, accurate determination of 
the individual planet parameters will be difficult due to the complexity of 
multiple planet lensing models.

In this paper, we propose a new channel to search for extra-solar systems 
composed of multiple planets via microlensing.  The method is based on the 
finding of \citet{han01b} that the lensing light curve anomalies induced by 
multiple planets are well approximated by the superposition of those of the 
single planet systems where the individual planet-primary pairs act as 
independent lens systems.  Then, if the source trajectories passes both of 
the outer deviation regions around the planetary caustics of the individual 
planets, one can identify the existence of the multiple planets.  We 
illustrate that the probability of successively detecting lensing light 
curve anomalies induced by two Jovian-mass planets located in the lensing 
zone through this channel will be substantial.  We discuss the advantages 
of the proposed method over the previous method of monitoring high 
magnification events.

\section{Basics of Multiple Planet Lensing}

If a source located at $\rvec_{s}$ on the projected plane of the sky is 
lensed by a lens system composed of $N$-point masses, where the individual 
components' masses and locations are $m_{i}$ and $\lvec_{i}$, the positions 
of the resulting images, $\rvec$, are obtained by solving the lens equation 
of the form
\begin{equation}
\rvec_{s} \ =\ 
\rvec \,-\, \theta_{\rm E}^{2} \
\sum_{i=1}^{N} {m_i\over m} \ 
{\rvec - \lvec_{i} \over |\rvec - \lvec_i|^2} \ ,
\end{equation}
where $m=\sum_{i}^{N} m_{i}$ is the total mass of the lens system and 
$\theta_{\rm E}$ is the angular Einstein ring radius.  The angular Einstein 
ring radius is related to the physical parameters of the lens system by
\begin{equation}
\theta_{\rm E}=\sqrt{{4Gm\over c^2}} 
	       \left( {1\over D_{ol}}-{1\over D_{os}}\right)^{1/2},
\end{equation}
where $D_{ol}$ and $D_{os}$ are the distances to the lens and the source 
from the observer, respectively.  Since the lensing process conserves the 
surface brightness of the source, the magnification of each image is simply 
given by the surface area ratio between the image and the unamplified source, 
and mathematically its value corresponds to the inverse of the Jacobian of 
the lens equation evaluated at the image position $\rvec_{j}$;
\begin{equation}
A_{j} \ =\ 
\left( {1\over |\det J|} \right)_{\rvec=\rvec_{j}} \ ;
\ \ \
\det J \ =\ 
\left\vert {\partial \rvec_{s} \over \partial \rvec} \right\vert \,.
\end{equation}
The total magnification is then given by the sum of the individual images' 
magnifications, i.e.\ $A_{\rm tot}=\sum_{j}^{N_I} A_{j}$, where $N_I$ is 
the total number of images.

For a single point-mass lens ($N=1$), there are two solutions for the lens 
equation ($N_I=2$) and the resulting total magnification is expressed in a 
simple analytical form of
\begin{equation}
A \ =\ 
\frac{u_{s}^{2}+2}{u_{s} \sqrt{u_{s}^2+4}} \ ,
\end{equation}
where $\uvec_{s}=(\rvec_{s}-\lvec)/\theta_{\rm E}$ is the dimensionless
lens-source separation vector normalized by $\theta_{\rm E}$.  For a 
rectilinear lens-source transverse motion, the separation vector is related 
to the single lensing parameters by
\begin{equation}
\uvec_{s} \ =\ 
\left({t-t_{0}\over t_{\rm E}}\right)\,\hat{\xvec}\,+\,\beta\,\hat{\yvec}\,
\end{equation}
where $t_{\rm E}$ represents the time required for the source to transit 
$\theta_{\rm E}$ (Einstein time scale), $\beta$ is the closest lens-source 
separation in units of $\theta_{\rm E}$ (impact parameter), $t_0$ is the 
time at that moment, and the unit vectors $\hat{\xvec}$ and $\hat{\yvec}$ 
are parallel with and normal to the direction of the relative lens-source 
transverse motion, respectively.

The lens system with planets is described by the formalism of multiple 
lensing with very low mass companions.  For a binary lens system ($N=2$), 
the lens equation becomes a fifth order polynomial in $r$ and the positions 
of the individual images are obtained by numerically solving the equation 
\citep{witt90}.  This yields three or five solutions depending on the source 
position with respect to the lenses.  The main new feature of a multiple lens 
system is the formation of caustics.  For a binary lens, the set of caustics 
form several disconnected closed curves.  The number and locations of the 
caustic curves are dependent on the separation between the planet and the 
primary lens.  If the separation is larger than $\theta_{\rm E}$, there exist 
two sets of caustics and one is the central caustic located near the primary 
lens and the other is the planetary caustic located away from the primary on 
the planet side with respect to the primary.  The lens system having a planet 
with a separation smaller than $\theta_{\rm E}$ also has a single central 
caustic, but has two planetary caustics located on the opposite side of the 
planet.  The planetary caustic(s) is (are) located within the Einstein ring 
when the projected separation between the planet and the primary (normalized 
by $\theta_{\rm E}$) is in the lensing zone of $0.618\leq b\leq 1.618$.  
Since the sizes of both the central and planetary caustics are maximized at 
around this separation, planet detection probability is also maximized for 
systems having planets located in this lensing zone \citep{gould92}.

For a lens system with triple lenses ($N=3$), the lens equation becomes a 
tenth order polynomial and it is still numerically solvable.  For these lens
systems, there are a maximum of ten images and a minimum of four images, 
and the number of images changes by a multiple of two as the source crosses a 
caustic.  Unlike the caustics of a binary lens system forming separate sets 
of closed curves, those of a triple lens system can exhibit self-intersection 
and nesting.

\begin{figure*}
\plotfiddle{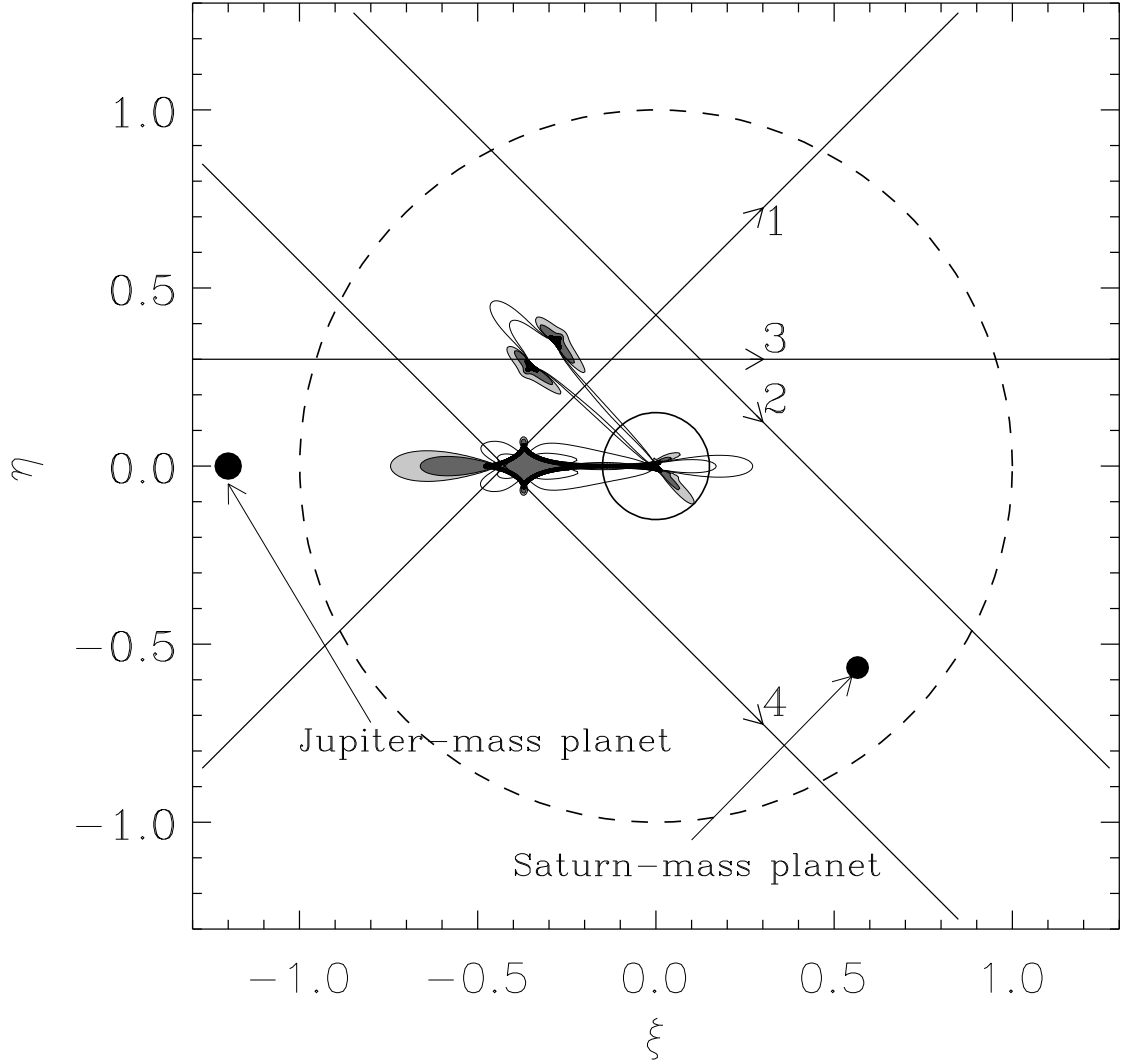}{0.0cm}{0}{110}{110}{-310}{-810}
\vskip13.0cm
\caption{
The map of magnification excess of an example lens system composed of two
planets with mass ratios of 0.003 and 0.001, which are comparable to those
of Jupiter-mass and Saturn-mass planets around a $\sim 0.3\ M_\odot$ star, 
respectively.  The coordinates $(\xi,\eta)$ are set so the the primary is 
at the origin and $\xi$ axis is parallel with the axis connecting the primary 
and the Jupiter-mass planet.  All lengths are scaled by the angular Einstein 
ring radius.  The projected separations (normalized by $\theta_{\rm E}$) of 
the individual planets from the host star are $b_J=1.2$ and $b_S=0.8$, 
respectively.  The angle between the the position vectors to the planets 
from the host star is $\alpha=135^\circ$.  Contours are drawn at the levels 
of $\epsilon=-0.1$, $-0.05$, 0.05, and 0.1 and the regions of positive 
deviations are distinguished by grey scales.  The small figures drawn by 
thick solid lines are the caustics and the dashed circle is the Einstein 
ring.  The region enclosed by  solid circle around the primary with a radius 
$u_s=0.15$ represents the central deviation region, which is excluded in our 
probability estimation of detecting both planets.  The straight lines with 
arrows represent the source trajectories of several example events for which 
one can detect both (the trajectory designated by a number `1'), one (`3' 
and `4'), and none (`2') of the planets, respectively.
}
\end{figure*}

For lens systems with even larger number of lenses ($N\geq 4$), numerically
solving the lens equation becomes very difficult.  However, one can still
obtain the magnification patterns of these lens systems by using the inverse
ray-shooting technique \citep{schneider86, kayser86, wambsganss90}.  In
this method, a large number of light rays are uniformly shot backwards
from the observer plane through the lens plane and then collected (binned) 
in the source plane.  Then, the magnification pattern is obtained by the 
ratio of the surface brightness (i.e., the number of rays per unit area) on 
the source plane to that on the observer plane.  Then the light curve 
resulting from a particular source trajectory corresponds to the 
one-dimensional cut through the constructed magnification pattern.  This 
technique has an advantage of allowing one to construct magnification 
patterns regardless of the number of lenses, but it has a disadvantage of 
requiring large computation time to obtain smooth magnification patterns.

\section{A New Channel}

Recently, \citet{han01b} showed that the light curve anomalies induced
by multiple planets are well approximated by the superposition of those 
of the single planet lens systems where the individual planet-primary 
pairs act as independent lens systems.  As pointed out by Gaudi et al.\ 
(1998), then, the anomaly pattern in the central region caused by one 
of the planets can be significantly affected by the existence of other 
planet(s) because the central deviation regions caused by the individual 
planets occur in this same region.  However, the anomaly pattern 
in the outer deviation region caused by each planet is hardly affected 
by other planet(s) because the outer deviation regions of the individual 
planets occur, in general, at different locations.

The fact that the anomaly patterns in the outer deviation regions are 
scarcely affected by other planet(s) provides a new channel to search for 
multiple planets and to determine their parameters.  In this channel, 
multiple planets are detected when the source trajectory passes both of the 
outer deviation regions around the planetary caustics of the individual 
planets.  The greatest advantage of this method over the method of monitoring 
high magnification events is that one can unambiguously identify the existence 
of multiple planets from the unique pattern of consecutive anomalies and 
accurately determine the planet parameters due to the applicability of much 
simpler single planet lensing models to the individual anomalies.

\section{Efficiency of the New Channel}

In this section, we illustrate that the efficiency of detecting multiple
planets through the new channel will be substantial.  To demonstrate this, 
we estimate the probability of successively detecting lensing light curve 
anomalies induced by multiple planets.  The probability is determined for 
a lens system having two Jovian-mass planets (i.e.\ Jupiter and Saturn) 
orbiting a host star with a mass of $\sim 0.3\ M_\odot$.  Then, the mass 
ratios of the planets (to that of the host star) are $q_J=0.003$ and $q_S
=0.001$, for the Jupiter-mass and the Saturn-mass planets, respectively.
If the lens system is located located at $D_{ol}\sim 6$ kpc (and with a 
source at $D_{os}\sim 8$ kpc) and the component planets have intrinsic 
orbital separations similar to those of Jupiter and Saturn of our solar 
system, the orbital separations (normalized by $\theta_{\rm E}$) of the 
planets correspond to $a_J\sim 2.7$ and $a_S\sim 5.0$, respectively.

To estimate the probability, we first construct maps of magnification 
excess.  The magnification excess is defined by
\begin{equation}
\epsilon = {A-A_0\over A_0},
\end{equation}
where $A$ and $A_0$ represent the magnifications expected with and without 
the planets, respectively.  Figure 1 shows the constructed map when the 
projected separations of the individual planets are $b_J=1.2$ and $b_S=0.8$ 
and the angle between the position vectors to the planets fro the host star 
(orientation angle) is $\alpha=135^\circ$.  We note that the projected 
separation, $b$, is related to the intrinsic orbital separation, $a$, by
\begin{equation}
b = a \sqrt{\cos^2\phi + \sin^2\phi\ \cos i},
\end{equation}
where $\phi$ is the phase angle and $i$ is the inclination of the orbital 
plane.

Once the map is constructed, the probability is estimated by computing the 
ratio of events whose source trajectories pass both of the outer deviation
regions induced by the individual planets (for example, the event resulting 
from the source trajectory designated by `1' in Fig.\ 1) during the 
measurements to the total number of trial events.  For given lens positions, 
the source trajectory orientations (with respect to the Jupiter-primary axis) 
and the impact parameter (with respect to the primary) of the trial events 
are randomly selected in the ranges of $0\leq\theta\leq 2\pi$ and $0\leq 
\beta\leq 1.0$, respectively.  Measurements for each event are assumed to 
be carried out during $-t_{\rm E}\leq t_{obs}\leq t_{\rm E}$ with a frequency 
of 10 times/night, which corresponds to that of the current microlensing 
followup observations \citep{rhie99, albrow98, bond01}.  We consider each 
planet is detected if excesses greater than a threshold value of 
$\epsilon_{th}=5\%$ are consecutively detected more than 5 times during the 
measurements.  Since we are interested in anomalies occurred only in the 
outer deviation regions, we do not count the detections of anomalies occurred 
in the central deviations region (the region enclosed by the solid circle 
around the primary with a radius $u_s=0.15$).  In addition, since identifying 
the existence of both planets and determining their parameters will be 
difficult if both of the planets' outer deviation regions are located at a 
similar place due to the resulting complexity of the interfered pattern of 
anomalies, we also do not count detections if the time interval between the 
anomalies induced by the individual anomalies is shorter than $0.1 t_{\rm E}$.

\begin{figure}[t]
\plotfiddle{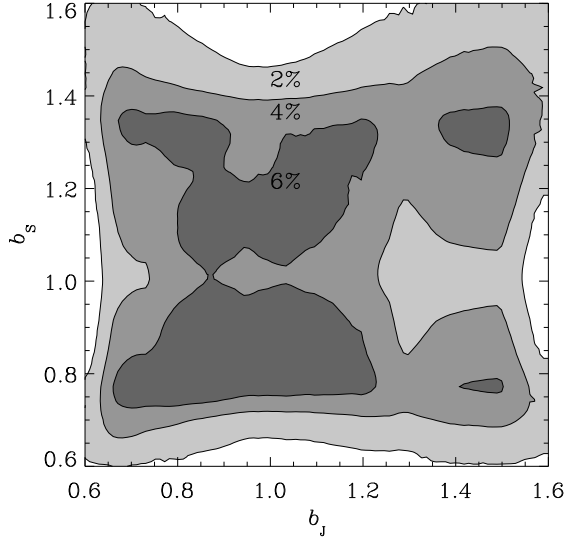}{0.0cm}{0}{55}{55}{-157}{-415}
\vskip6.8cm
\caption{
The distribution of the absolute probabilities to detect both planets of
the example lens system with two Jovian-mass planets as functions of the
projected separations of the individual planets from the host star, 
$P_{\rm abs}(b_J,b_S)$.  The observational conditions and the detection 
criteria are described in \S\ 4.
}
\end{figure}

In Figure 2, we present the distribution of the {\it absolute} probabilities 
of detecting both planets as functions of their projected separations, 
$P_{abs}(b_J,b_S)$.  The presented probabilities are the values averaged 
over the random orientation angles in the range of $0\leq\alpha\leq 2\pi$.  
In Figure 3, we also present the distribution of {\it conditional} 
probabilities to successively detect the second planet under the condition 
that the first planet is detected, $P_{cond}(b_J,b_S)$.  One finds that if 
the two planets are located in the lensing zone, detecting both of them
is possible with absolute probabilities of $P_{abs}\gtrsim 2\%$.  One also 
finds that once a planet is detected, the probabilities to successively 
detect the second one are $P_{cond}\gtrsim 10\%$.

\begin{figure}[t]
\plotfiddle{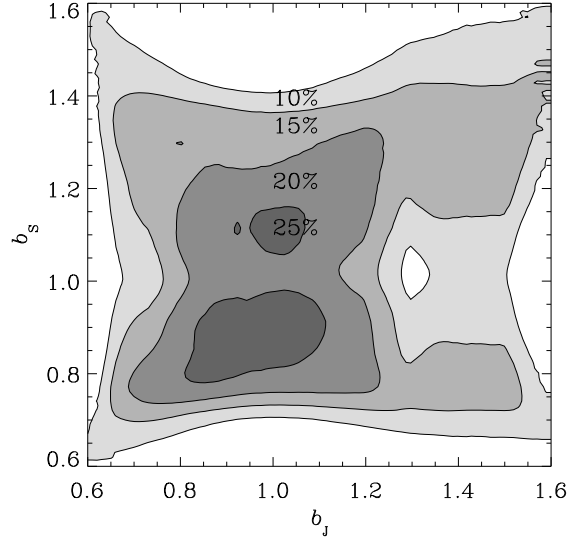}{0.0cm}{0}{55}{55}{-157}{-415}
\vskip6.5cm
\caption{
The distribution of the conditional probabilities to successively detect
the second planet under the condition that the first planet is detected,
$P_{\rm abs}(b_J,b_S)$.
}
\end{figure}

\section{Conclusion}
We propose a new channel of detecting extra-solar systems composed of having 
multiple planets by using microlensing.  In this method, multiple planets 
are detected when the source trajectory passes both of the outer regions 
of deviations induced by the individual planets.  From the estimation of 
the probabilities to detect both planets of an example Galactic lens system 
composed of two Jovian-mass planets around a star with $\sim 0.3\ M_\odot$,
we find that if they are located within the lensing zone, both planets can 
be detected with a non-negligible absolute probabilities ($P_{abs}\gtrsim 
2\%$) and a substantial conditional probabilities ($P_{cond} \gtrsim 10\%$) 
of successively detecting the second planet under the condition that the 
first planet is detected.  The proposed method has an important advantage 
of allowing one to accurately determine the planet parameters because the 
light curve anomalies induced by the individual planets can be well described 
by simple single planet lensing models.

\acknowledgments
This work was supported by a grant (2000-015-DP0449) from Korea Research 
Foundation (KRF).

\end{document}